\documentclass[aps, showpacs, twocolumn]{revtex4}
\usepackage{amsmath,amssymb}

\usepackage{tabularx}

\usepackage{epsfig}
\usepackage{graphicx}

\begin{document}

\title{Detectable primordial non-gaussianities and gravitational waves in k-inflation}

\author{Grigoris Panotopoulos\footnote{grigoris@theorie.physik.uni-muenchen.de}}

\date{\today}


\address{ASC, Department of Physics LMU, Theresienstr. 37, 80333 Munich, Germany}


\begin{abstract}
An inflationary single field model with a non-trivial kinetic term for the inflaton is discussed. It is shown that it is possible to have large primordial non-gaussianities and large tensor-to-scalar ratio in a simple concrete model with just a scalar field and a generalized kinetic term for the inflaton field. This is potentially interesting in the prospect of new forthcoming observations.
\end{abstract}


\maketitle

\section{Introduction}

The inflationary phase of the early universe~\cite{inflation} is a very important ingredient in modern cosmology. The reason for this is twofold. First, inflation solves the longstanding problems of standard hot big-bang cosmology, namely the horizon, flatness and monopole problems. Furthermore, it produces the primordial perturbations for the structure that we observe today. The recent spectacular CMB data from the WMAP satellite~\cite{wmap} have strengthen the inflationary idea, since the observations indicate an \emph{almost} scale-free spectrum of Gaussian adiabatic density fluctuations, just as predicted by simple models of inflation.

In the prospect of future CMB experiments, like~\cite{planck}, constructing inflationary models which predict detectable primordial gravitational waves (tensor-to-scalar ratio) and non-gaussianities has attracted a lot of interest lately. As far as tensor-to-scalar ratio is concerned, at present the limit is $r \leq 0.55$ and it easy to build inflationary models that predict negligible tensor perturbations. However, it would be interesting to have at our disposal models which predict a tensor-to-scalar ratio at the future detection limit, $r \geq 0.001-0.01$. In fact, as it is known that current superstring inflationary models (see e.g.~\cite{reviews}) predict negligible tensor perturbations~\cite{KL} (see however~\cite{krause}), this would be challenging for testing string theory. On the other hand, it is known that single field inflationary models with a minimally coupled scalar field predict primordial non-gaussianities at a non-detectable level~\cite{maldacena}, since the non-linear parameter $f_{NL} \ll 1$. However, the PLANCK satellite is expected to detect the non-linear parameter if $|f_{NL}| \geq 5$~\cite{komatsu1}. This is the reason why great effort is devoted to the construction of inflationary models which predict large non-gaussianities~\cite{models, DBI}.

In the present work we show that a simple model with a single scalar field and with a non-trivial kinetic term can lead to detectable tensor-to-scalar ratio $r$ and non-linear parameter $|f_{NL}|$. Scalar field models with a non-canonical kinetic term have been discussed in k-inflation~\cite{kinflation}, in which inflation is not due to the potential but rather to the kinetic term in the lagrangian, and in k-essence models~\cite{kessence}, which are designed to address the issue why the cosmic acceleration has recently begun.  A previous work~\cite{slava} is closely related to ours. In this, the authors have shown that within the same class of models it is possible to have enhanced production of gravitational waves without conflict with observations. However, non-gaussianities are not-discussed in this paper and in fact, as we will show below, the particular model~\cite{slava} predicts negligible $f_{NL}$. The reason for that is that in the model of~\cite{slava} the sound speed $c_s^2 > 1$. In the present work however we study a different model in which the sound speed can be very low, $c_s^2 \ll 1$, and therefore it is possible to have large $r$ and $|f_{NL}|$ at the same time.

Our work is organized as follows. The article consists of four sections, of which this introduction is the first. We present the basic formula in the second section and then we discuss slow-roll inflation in a simple model in section 3. Finally the fourth section is devoted to some conclusions.

\section{Dynamics of a generalized scalar field}

Our framework is four-dimensional General Relativity coupled to a single scalar field with a general lagrangian $\mathcal{L}(\phi, X)$, where $\phi$ is the scalar field and $X=(\partial \phi)^2/2$ is the standard kinetic term. Therefore, our model is described
by the action
\begin{equation}
S=\int d^4x \sqrt{-g} \: \left (-\frac{R}{16 \pi}+\mathcal{L}(\phi, X) \right)
\end{equation}
and we have set Newton's constant $G=1$. The energy-momentum tensor for the scalar field is given by
\begin{equation}
T_{\mu \nu}^{(\phi)}=\mathcal{L}_{,X} \partial_{\mu} \phi \partial_{\nu} \phi-\mathcal{L} g_{\mu \nu}
\end{equation}
where $,X$ denotes differentiation with respect to $X$. We can recast this energy-momentum tensor into the form of the energy-momentum tensor for a perfect fluid
\begin{equation}
T_{\mu \nu}^{(p.f)}=(\epsilon + p) u_{\mu} u_{\nu}-p g_{\mu \nu}
\end{equation}
where $\epsilon, p$ are the energy density and the pressure of the fluid respectively.
The hydrodynamical quantities $\epsilon, p, u_{\mu}$ are given in terms of $\phi, X$ as follows
\begin{eqnarray}
p & = & \mathcal{L} \\
\epsilon & = & 2Xp_{,X}-p \\
u_{\mu} & = & \frac{\partial_{\mu} \phi}{\sqrt{2X}}
\end{eqnarray}
In the case in which the lagrangian of the scalar field is of the form $\mathcal{L}=K(X)-V(\phi)$, the energy density is given by
\begin{equation}
\epsilon=2XK_{,X}-K+V
\end{equation}
Therefore, the equations of motion for our system are just the first Friedmann equation and the equation for energy conservation
\begin{eqnarray}
H^2 & = & \frac{8 \pi}{3} \: \epsilon \\
\dot{\epsilon} & = & -3H(\epsilon + p)
\end{eqnarray}
where $H=\dot{a}/a$ is the Hubble parameter, $a$ is the scale factor and the overdot denotes differentiation with respect to cosmic time. Defining the sound speed of the scalar field
\begin{equation}
c_s^2 \equiv \frac{p_{,X}}{\epsilon_{,X}}=\left ( 1+2X \: \frac{p_{,XX}}{p_{,X}} \right )^{-1}
\end{equation}
the equation for energy conservation takes the form
\begin{equation}
\ddot{\phi}+3c_s^2H\dot{\phi}+\frac{\epsilon_{,\phi}}{\epsilon_{,X}}=0
\end{equation}
which generalizes the usual Klein-Gordon equation of a canonical scalar field in a Friedmann-Robertson-Walker background.

\section{Slow-roll inflation in a simple model}

During the slow-roll phase the inflationary dynamics can be quantified in terms of the three parameters
\begin{eqnarray}
e & = & \frac{2 M_p^2}{\gamma} \: \left ( \frac{H'}{H} \right )^2 \\
\eta & = & \frac{2 M_p^2}{\gamma} \: \frac{H''}{H} \\
s & = & \frac{2 M_p^2}{\gamma} \: \frac{H'}{H} \: \frac{\gamma'}{\gamma}
\end{eqnarray}
where $\gamma=1/c_s$, $s \equiv \dot{c_s}/(Hc_s)$, $M_p$ is the reduced Planck mass (in the units we are using here $M_p^2=1/(8 \pi)$) and a prime denotes differentiation with respect to the scalar field. We assume that $(\epsilon, |\eta|, |s|) \ll 1$ during the slow-roll inflationary phase. Furthermore, assuming that $\mathcal{L}=K(X)-V(\phi)$ the equations of motion in the slow-roll approximation take the form
\begin{eqnarray}
H^2 & = & \frac{8 \pi}{3} \: V \\
0 & = & 3c_s^2H\dot{\phi}+\frac{V_{,\phi}}{\epsilon_{,X}}
\end{eqnarray}
since we ignore $\ddot{\phi}$ in the equation of motion for the scalar field, and $V$ is the dominant term in the expression for the energy density. Upon considering a concrete model in which
\begin{eqnarray}
V(\phi) & = & \frac{1}{2} \: m^2 \: \phi^2 \\
K(X) & = & \alpha X^{\beta}
\end{eqnarray}
we find that the speed of sound is a constant and given by
\begin{equation}
c_s^2=\frac{1}{2 \beta -1}
\end{equation}
and it can be very low for large $\beta$, $\beta \gg 1$. In addition, the equations of motion now take the form
\begin{eqnarray}
H & = & \sqrt{\frac{4 \pi}{3}} m \phi \\
0 & = & 3p_{,X}H\dot{\phi}+m^2 \phi
\end{eqnarray}
Notice that in this model $s=0=\eta$ and the end of inflation is determined by the condition $e_{end}=1$.

Finally we define the number of e-folding
\begin{equation}
N=\int dt H
\end{equation}
as well as the quantities that connect the theoretical model to observations. These are the spectral index and the amplitude of perturbations both for scalar and tensor perturbations. The perturbations in k-inflation in linear order have been computed in~\cite{perturbations}. In linear perturbation theory the scalar and tensor perturbations evolve independently from each other and therefore scalar perturbations cannot produce gravitational waves. However this is possible if second order effects are taken into account~\cite{2order}. Since k-inflation generalizes the usual inflation with a canonical scalar field, we expect that we could study the generation of gravitational waves by second order effects in k-inflation generalizing the results of~\cite{2order}. This however lies beyond the scope of the present work. The indices and amplitudes of scalar and tensor perturbations are given by
\begin{eqnarray}
P_S^2 & = & \frac{H^4}{4 \pi^2 \dot{\phi}^2} \\
P_T^2 & = & \frac{2 H^2}{\pi^2 M_p^2} \\
n_T & = & -2 e \\
n_S & = & 1+2 \eta-4e-2s
\end{eqnarray}
The tensor-to-scalar ratio is defined by $r \equiv P_T^2/P_S^2$ and is given by
\begin{equation}
r=\frac{P_T^2}{P_S^2}=16 c_s e
\end{equation}
In standard inflation ($c_s=1$) the tensor spectral index and the tensor-to-scalar ratio are not independent, and one has the so-called "consistency relation"~\cite{LL}, $r=-8n_T$, which in principle can be tested against observations. However, in k-inflation this relation does not hold any more. Instead of that we have $r=-8 c_s n_T$. Thus, at least in principle, kinetic inflation is phenomenologically distinguishable from standard inflation~\cite{perturbations}. If in fact $c_s > 1$ the production of gravitational waves is substantially larger than expected in standard inflationary models. This was pointed out in~\cite{slava}.

As in~\cite{slava}, the equation of motion for the scalar field in the slow-roll approximation can be easily solved and the solution corresponds to a constant kinetic term $X$. However, this is not important for what we would like to point out. Furthermore, we are not going to concern ourselves with the amplitude of scalar perturbations or the scalar spectral index, since in the present work our main interest is being on the tensor-to-scalar ratio and the non-gaussianity. We just remark in passing that the observable quantities, $P_s, n_s$ etc, depend on the three model parameters $\alpha, \beta, m$ through the slow-roll parameters. For a given value of $\beta$ or $c_s^2$ we can determine the values of $m, \alpha$ by requiring that we must reproduce the observable values of $P_s$ and $n_s$. Then we can compute the rest of the observable quantities, namely the tensor-to-scalar ratio $r$ and the non-linear parameter $f_{NL}$, to which we now turn.

Let us now focus on the primordial non-gaussianity. A rough measure of the non-gaussianities is provided by the non-linear parameter $f_{NL}$
\begin{equation}
\Phi(x)=\Phi_L(x)+f_{NL} (\Phi_L(x)^2-\langle \Phi_L(x)^2 \rangle)
\end{equation}
where $\Phi(x)$ is the gravitational potential which sources the temperature anisotropies in the CMB~\cite{komatsu2}. In our model the parameter $f_{NL}$ in leading order is given by~\cite{3-point}
\begin{equation}
f_{NL}=-0.28 (1-\frac{1}{c_s^2})
\end{equation}
According to~\cite{3-point}, $f_{NL}$ is given by the expression
\begin{equation}
f_{NL} \simeq -0.28 u+0.02 s \frac{e}{\epsilon}-1.53e-0.42 \eta
\end{equation}
where $u=1-1/c_s^2.$ For usual inflation with a canonical scalar field, $c_s=1, s=0=u$ and therefore $f_{NL} \sim 0.01$ as we mentioned before. In DBI inflation~\cite{DBI}  $s,u$ are different than zero and the sound speed $c_s^2 < 1$. In a generic k-essence model $s,u$ are also different than zero, but the sound speed can be either larger than unity, $c_s^2 > 1$ or lower than unity, $c_s^2 < 1$. In our model in the present work $c_s^2=$ constant, $s=0$, $c_s^2 \ll 1$ and $u$ different than zero.

It easy to see that for a large sound speed, $c^2 \gg 1$, the non-linear parameter $f_{NL} \simeq -0.28$, which is just one order of magnitude compared to the canonical scalar field case, $f_{NL} \sim \epsilon \sim 0.01$, while for a tiny sound speed, $c_s^2 \ll 1$, the non-linear parameter $f_{NL} \simeq 0.28/c_s^2$. In this case $f_{NL}$ can be much greater and therefore within the detectable range. For example, let us consider a case in which $e=1/100$, $\beta=13$ and $c_s^2=1/25$. Then we obtain for $r$ and $f_{NL}$ the values
\begin{eqnarray}
r & = & 0.032 \\
f_{NL} & = & 6.72
\end{eqnarray}
either of which is within the detectable range. In the figure below we show $f_{NL}$ as a function of $r$ and with the two horizontal lines we show the region that corresponds to a detectable range both for $r$ and $f_{NL}$.

\begin{figure}[h!]
\centering
\begin{tabular}{cc}
\includegraphics*[width=260pt, height=260pt]{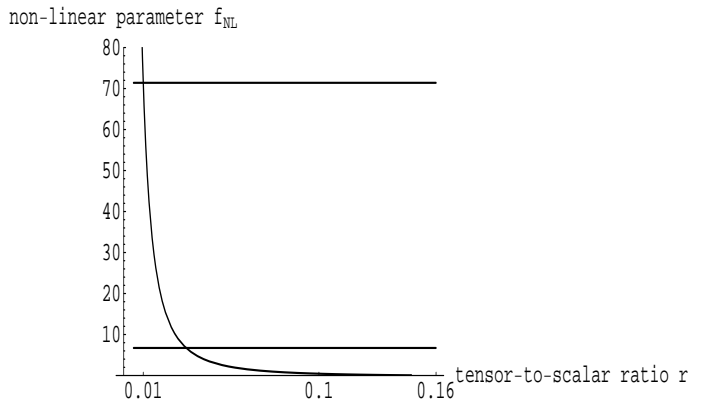}&%
\end{tabular}
 \caption{Non-linear parameter $f_{NL}$ as a function of the tensor-to-scalar ratio $r$. The horizontal lines correspond to $f_{NL}=6.72=f_{NL}(r=0.032)$ and $f_{NL}=71.4=f_{NL}(r=0.01)$.}
\end{figure}

In fact our discussion shows that even the form of the potential $V(\phi)=m^2 \phi^2/2$ is not important. What is important here is the form of the kinetic term, $X=\alpha X^{\beta}$. The key quantity in the present work is the sound speed and the two facts that it is constant and that is may become low enough so that we can have a large non-linear parameter $f_{NL}$, but not a negligible tensor-to-scalar ratio $r$. Therefore we expect that in our simple inflationary model it is possible to obtain both $r$ and $f_{NL}$ in the detectable range independently of the form of the inflaton potential $V(\phi)$.

A final remark regarding the sound speed is in order here. It is known that for a canonical scalar field the speed of sound equals unity. However in models with a general scalar field the speed of sound can be different from unity and in fact it can even be $c_s^2>1$. For a recent discussion on superluminal fields and causality one can see~\cite{causality}. In our work we have considered a case in which the speed of sound is very small, $c_s^2 \ll 1$. In the past k-essence models received some criticism because of possible superluminal propagation of the field fluctuations~\cite{durrer}. However it was shown recently that causal paradoxes do not arise in this kind of models~\cite{mukhanov}.

\section{Conclusions}

In the present work we have studied slow-roll inflation with a single scalar field and non-trivial kinetic term for the inflaton. In particular, we have considered a simple concrete model in which the lagrangian for the generalized scalar field is of the form $\mathcal{L}=K(X)-V(\phi)$, where $K(X)=\alpha X^{\beta}$ is the non-trivial kinetic term and $V(\phi)=m^2 \phi^2/2$ is the inflaton potential. In this model the speed of sound is constant and it can be very low for $\beta \gg 1$. We have shown that in this simple model it is possible to have large tensor-to-scalar ratio and large primordial non-gaussianities. This is a potentially interesting result in the prospect of new forthcoming observations. In fact our main result is probably true for any form of the inflaton potential $V(\phi)$, as it turns out that the crucial point in our discussion is the form of the non-trivial kinetic term.

\section*{Acknowlegements}

We would like to thank the anonymous reviewer for his/her valuable comments and suggestions. This work was supported by project "Particle Cosmology".

\end{document}